\documentclass[conference]{IEEEtran}
\IEEEoverridecommandlockouts
\usepackage{cite}
\usepackage{amsmath,amssymb,amsfonts}
\usepackage{algorithmic}
\usepackage{graphicx}
\usepackage{caption}
\usepackage{braket}
\usepackage{nth}
\usepackage{subcaption}
\usepackage{algorithm}

\usepackage{textcomp}
\usepackage{xcolor}
\def\BibTeX{{\rm B\kern-.05em{\sc i\kern-.025em b}\kern-.08em
    T\kern-.1667em\lower.7ex\hbox{E}\kern-.125emX}}
\begin{document}

\title{Amplitude-Aware Lossy Compression for Quantum Circuit Simulation\\}

\author{\IEEEauthorblockN{Xin-Chuan Wu}
\IEEEauthorblockA{\textit{Department of Computer Science} \\
\textit{University of Chicago}\\
Chicago, USA \\
xinchuan@uchicago.edu}
\and
\IEEEauthorblockN{Sheng Di}
\IEEEauthorblockA{\textit{Mathematics and Computer Science} \\
\textit{Argonne National Laboratory}\\
Lemont, USA \\
sdi1@anl.gov}
\and
\IEEEauthorblockN{Franck Cappello}
\IEEEauthorblockA{\textit{Mathematics and Computer Science} \\
\textit{Argonne National Laboratory}\\
Lemont, USA \\
cappello@mcs.anl.gov}
\and
\IEEEauthorblockN{Hal Finkel}
\IEEEauthorblockA{\textit{Argonne Leadership Computing Facility} \\
\textit{Argonne National Laboratory}\\
Lemont, USA \\
hfinkel@anl.gov}
\and
\IEEEauthorblockN{Yuri Alexeev}
\IEEEauthorblockA{\textit{Computational Science} \\
\textit{Argonne National Laboratory}\\
Lemont, USA \\
yuri@anl.gov}
\and
\IEEEauthorblockN{Frederic T. Chong}
\IEEEauthorblockA{\textit{Department of Computer Science} \\
\textit{University of Chicago}\\
Chicago, USA \\
chong@cs.uchicago.edu}
}

\maketitle

\begin{abstract}
Classical simulation of quantum circuits is crucial for evaluating and validating the design of new quantum algorithms. However, the number of quantum state amplitudes increases exponentially with the number of qubits, leading to the exponential growth of the memory requirement for the simulations. In this paper, we present a new data reduction technique to reduce the memory requirement of quantum circuit simulations. We apply our amplitude-aware lossy compression technique to the quantum state amplitude vector to trade the computation time and fidelity for memory space. The experimental results show that our simulator only needs 1/16 of the original memory requirement to simulate Quantum Fourier Transform circuits with 99.95\% fidelity. The reduction amount of memory requirement suggests that we could increase 4 qubits in the quantum circuit simulation comparing to the simulation without our technique. Additionally, for some specific circuits, like Grover's search, we could increase the simulation size by 18 qubits.
\end{abstract}


 \section{Introduction}\label{sec:intro}

Classical simulation of quantum circuits is crucial for better understanding the operations and behaviors of quantum computers. Such simulations allow researchers and developers to evaluate the complexity of new quantum algorithms and validate the design of quantum circuits. Previous studies have provided different simulation techniques, such as full amplitude-vector update \cite{de2007massively, smelyanskiy2016qhipster, haner20170}, Feynman paths \cite{bernstein1997quantum}, and tensor network contractions \cite{markov2008simulating, pednault2017breaking, boixo2017simulation, chen201864, chen2018classical} to perform the simulation of quantum circuits. Full amplitude-vector update simulations provide all the amplitudes of the quantum state in detail, and hence it is the best tool for quantum algorithm debugging, development, and validation. Full amplitude-vector update simulators generally use complex double precision amplitudes to represent the state of the quantum systems. Given $n$ quantum bits (qubits), we need $2^n$ amplitudes to describe the quantum system \cite{nielsen2002quantum}. As a result, the size of the state vector is $2^{n+4}$ bytes. Since the number of quantum state amplitudes grows exponentially with the number of qubits in the system, the size of the quantum circuits simulation is restricted by the memory capacity of the classical computing system. For example, to store the full quantum state of a 45-qubit system, the memory requirement is 0.5 petabytes. Table \ref{tab:supercomputers} shows examples of several supercomputers and the largest quantum system they can simulate theoretically. Circuits with more than 49-qubit system would require too much memory to simulate directly.

\begin{table}
\caption{\textbf{Examples of supercomputers and the largest quantum system they can simulate.}} 
\label{tab:supercomputers}
\begin{center}
  \begin{tabular}{ l  c  c }
    \hline
    System & Memory (PB) & Max Qubits\\ \hline\hline
    TACC Stampede & 0.192 & 43\\ \hline
    Titan & 0.71 & 45\\ \hline
    Theta & 0.8 & 45\\ \hline
    K computer & 1.4 & 46\\ \hline
    Exascale & 4-10 & 48-49\\
    \hline
  \end{tabular}
\end{center}
\end{table}

Since the simulation size is restricted by the memory capacity of the classical computing systems, our goal is to trade the computation time for the memory space. By compressing the state vector, we can reduce the memory requirement for the quantum circuit simulation. In other words, we are able to simulate a larger quantum system by using the same memory capacity. Integrating data compression into the simulation causes certain performance overhead, while we are able to obtain larger simulation scale that cannot be reached before with the limited memory capacity. 

With compression techniques, the simulation size is determined by the quantum state vector's compression ratio. In general, lossy compression algorithms achieve significantly higher compression ratios than lossless compressors, while introducing errors to a certain extent. To minimize the error propagation and guarantee high fidelity results, we develop the \emph{Amplitude-Aware Lossy Compression} (AALC) technique by leveraging both Zstandard lossless compressor \cite{zstd} and SZ lossy compressor \cite{xin2018, tao2017significantly, di2016fast}. AALC is designed for quantum circuit simulators to trade data accuracy for data reduction. This technique can preserve the most important quantum information.

To evaluate our approach on a working simulator, we implement AALC on Intel-QS, a quantum circuit simulator developed by Intel \cite{smelyanskiy2016qhipster}. Intel-QS is a distributed high performance quantum circuit simulator, which uses full state amplitude-vector update to complete the simulation. Thus, there is no circuit depth limitation in our simulation framework\footnote{In general, most of the supercomputing systems have a 24-hour wall-time limitation. This run-time constraint puts a circuit depth limitation on the simulation. However, one can save the state vector before terminating the job, and then resume the task by loading the state vector in the next job submission.}. 

Our AALC approach integrates knowledge of quantum computation and data compression techniques to reduce the memory requirement of quantum circuit simulation. The main contributions of our work are:

\begin{itemize}
\item We identify the feasibility of applying data compression techniques to quantum circuit simulator.
\item We establish the AALC approach to leverage lossless and lossy compressor in quantum state amplitude vector to reduce the memory requirement for storing the data.
\item We present evaluation results based on AALC technique, and provide mathematical proofs that our technique can scale the simulation size beyond 49 qubits. 
\item We implement AALC for amplitude vectors in Intel-QS and show that there is a good data reduction with high quantum state fidelity.
\end{itemize}

Our paper is organized as follows. In Section~\ref{background}, we introduce background of  quantum circuit simulation. In Section~\ref{aalc}, we describe our AALC methodology. In Section~\ref{results}, we present the evaluation of our technique in Intel-QS. We discuss and point out the future directions in Section~\ref{future}.

\section{Background}\label{background}

\begin{figure}[!t]
\centering
\captionsetup{justification=centering}
\includegraphics[width=0.48\textwidth,keepaspectratio]{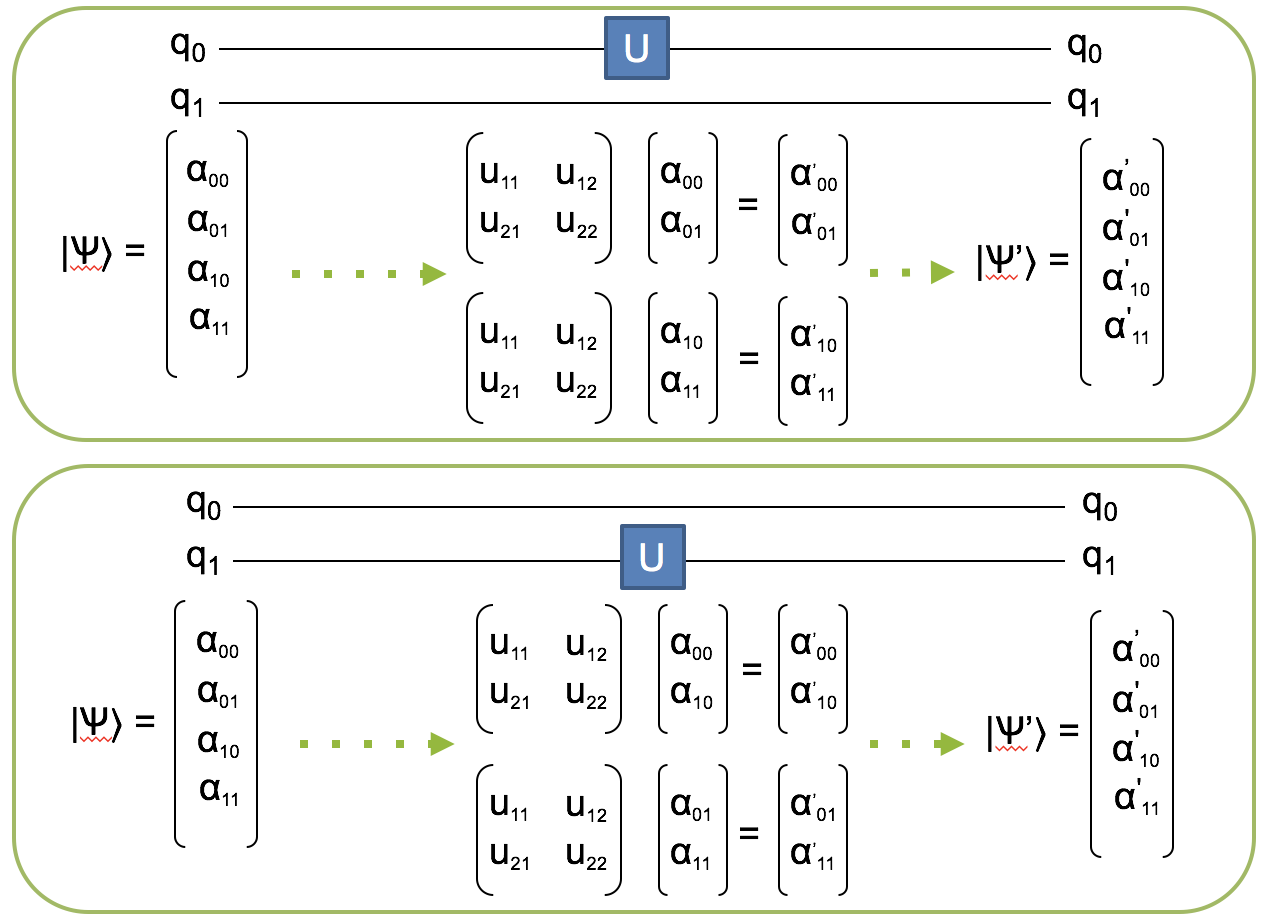}
\caption{Example of two-qubit quantum state with single-qubit gate operations.}
\label{fig:apply}
\end{figure}

A qubit is a two-level quantum systems, and the state $\ket{\psi}$ can be expressed as
$$\ket{\psi} = \alpha_0\ket{0} + \alpha_1\ket{1} $$
where $\alpha_0$ and $\alpha_1$ are complex amplitudes, and $|\alpha_0|^2 + |\alpha_1|^2 = 1$. $\ket{0}$ and $\ket{1}$ are two computational orthonormal basis states. The quantum state can also be represented as
$$\ket{\psi} = \alpha_0\begin{bmatrix}1\\0\end{bmatrix} + \alpha_1\begin{bmatrix}0\\1\end{bmatrix} = \begin{bmatrix}\alpha_0\\\alpha_1\end{bmatrix}.$$
The state of a two-qubit system can be described by using 4 amplitudes.
$$\ket{\psi} = \alpha_{00}\ket{00} + \alpha_{01}\ket{01} + \alpha_{10}\ket{10} + \alpha_{11}\ket{11} = \begin{bmatrix}\alpha_{00}\\\alpha_{01}\\\alpha_{10}\\\alpha_{11}\end{bmatrix}   .$$

More generally, the state of an n-qubit quantum system can be represented by using $2^n$ amplitudes.
$$\ket{\psi} = \alpha_{0...00}\ket{0...00} + \alpha_{0...01}\ket{0...01} + ... + \alpha_{1...11}\ket{1...11}.$$
In classical simulation of quantum computation, each amplitude is a double-precision complex number, which is a 16-byte data type. Thus, storing the full state vector of an n-qubit quantum system require $2^{n+4}$ bytes.

General single-qubit gates and two-qubit controlled gates are known to be universal \cite{divincenzo1995two}. Applying a quantum single-qubit gate $U$ to the $k\text{-th}$ qubit can be represented by a unitary transformation
$$A = I \otimes I \otimes ... \otimes U \otimes ... \otimes I \otimes I$$
where $I$ is $2\times 2$ identity matrix, and U is $2\times 2$ unitary matrix,
$$U = \begin{bmatrix}u_{11} & u_{12}\\u_{21} & u_{22}\end{bmatrix}.$$
However, we do not need to build the entire unitary matrix $A$ to perform the gate operation. Figure~\ref{fig:apply} shows an example of applying a single-qubit gate to a two-qubit system. Applying a gate $U$ to the first qubit is equivalent to applying the $2\times 2$ unitary matrix to every pair of amplitudes, whose subscript indices have 0 and 1 in the first bit, and all remaining bits are the same. In the same way, performing a single-qubit gate to the second qubit is to apply the unitary to every pair of amplitudes whose subscript indices differ in the second bit. More extensively, applying a single-qubit gate to the $k\text{-th}$ qubit of $n$-qubit quantum system is to apply the unitary to every pair of amplitudes whose subscript indices have 0 and 1 in the $k\text{-th}$ bit, while all other bits remain the same.
$$\begin{bmatrix}\alpha'_{*...*0_k*...*}\\\alpha'_{*...*1_k*...*}\end{bmatrix} = \begin{bmatrix}u_{11} & u_{12}\\u_{21} & u_{22}\end{bmatrix}\begin{bmatrix}\alpha_{*...*0_k*...*}\\\alpha_{*...*1_k*...*}\end{bmatrix} $$

As for a generalized two-qubit controlled gate, the unitary is applied to a target qubit $t$, if the control qubit $c$ is set to $\ket{1}$; otherwise $t\text{-th}$ is unmodified.
$$\begin{bmatrix}\alpha'_{*1_c...*0_t*...*}\\\alpha'_{*1_c...*1_t*...*}\end{bmatrix} = \begin{bmatrix}u_{11} & u_{12}\\u_{21} & u_{22}\end{bmatrix}\begin{bmatrix}\alpha_{*1_c...*0_t*...*}\\\alpha_{*1_c...*1_t*...*}\end{bmatrix} $$

In quantum information theory, \emph{fidelity} is a measure of distance between two quantum states. In this paper, we only consider pure states, so the definition of fidelity reduces to 
$$F(\rho, \sigma) = \braket{\psi_\rho|\psi_\sigma}.$$
For any $\rho$ and $\sigma$, $0 \leq F(\rho, \sigma) \leq 1$, and $F(\rho, \rho) = 1$.

\section{Amplitude-Aware Lossy Compression} \label{aalc}

We analyze the operations of the quantum circuit simulation, and notice that the gate operation can be performed separately pair by pair as described in Section~\ref{background}. This finding allow us to divide the whole state vector into several strides $(s_1, s_2, ..., s_n)$, and all the strides are stored in the compressed format on the memory, so that we can reduce the memory requirement for storing the state vector. The pseudo-code of the simulation process is shown in Algorithm~\ref{algo:operations}. The simulation of a quantum program is processed gate by gate. When a gate is applied to the quantum state, only the stride, $s_j$, under processing can be decompressed, and the unitary computation is performed on the decompressed stride, and then the stride is compressed. This is a complete operation cycle for a stride. After a stride is finished, we process the next stride.

\begin{algorithm}[htbp]
 \caption{Quantum state vector stride update.}
 \label{algo:operations}
\begin{algorithmic}[1]
 \FOR {gate $U_i$ in the program}
   \FOR{stride $s_j$ in the state vector}
   \STATE decompress($s_j$)\\
   \STATE gateComputation($U_i, s_j$)\\
   \STATE compress($s_j$)\\
  \ENDFOR
  \ENDFOR
 
\end{algorithmic}
\end{algorithm}

\begin{figure}[!t]
\centering
\captionsetup{justification=centering}
\includegraphics[width=0.48\textwidth,keepaspectratio]{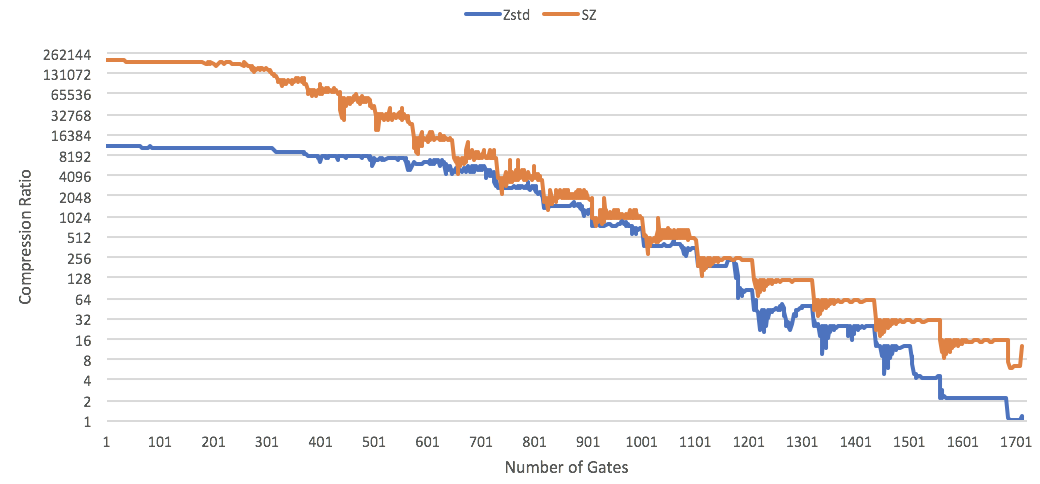}
\caption{Compression ratio results of QFT quantum circuit simulation.}
\label{fig:ratio}
\end{figure}

The straightforward idea is to apply lossless compression to the state vector. We leverage Zstandard (zstd) library \cite{zstd} as the lossless compression technique, and get a good compression ratio in the beginning of a quantum program simulation because most of the amplitudes in the initial state vector are zero, which is good for data compression. However, when a high complexity quantum program runs on the simulator, as more gate operations are performed, more and more non-zero amplitudes are generated in the state vector. As a result, the compression ratio is low in the end. Figure~\ref{fig:ratio} shows the compression ratio results from the Quantum Fourier Transform (QFT) circuit simulation, which consists of 26 qubits and 1710 quantum gates. In the first 400 gates, the compression ratio of the state vector can achieve 8192. However, after 1600 gates, the compression ratio is less than 2, and the minimal compression ratio is only 1.02. The minimal compression ratio is the most important metric to evaluate the memory requirement of the quantum circuit simulation. Thus, we have to improve the minimal compression ratio.

To improve the compression ratio, we introduce SZ lossy data compression technique \cite{xin2018, tao2017significantly, di2016fast} into our simulation framework. SZ allows users to set an error bound, $\delta$, for the compression. The decompressed data $D'_i$ might differ from the original data $D'_i$ but must be in the range $[D_i - \delta, D_i + \delta]$. Since lossy compression introduces compression errors into the state vector, the sum of the square of the amplitudes is no longer to be 1. To reduce the error propagation, the state vector is normalized before the gate computation.

We carefully examine the impact of compression errors on the simulation result, and trade the data accuracy for the data reduction. As shown in Figure~\ref{fig:ratio}, the minimum compression ratio with SZ lossy compressor is 5.68 and the state fidelity is 99.34\%. In addition, we can increase the error bound to get 8.02 compression ratio with 46.92\% state fidelity. However, the fidelity should be improved so that the simulation results can be useful for quantum algorithm development.

Our Amplitude-Aware Lossy Compression (AALC) technique is designed for further optimizing the minimal compression ratio and the state fidelity. The improved version of the simulation framework with AALC is shown in Algorithm~\ref{algo:aalc}. Instead of using a fixed error bound, AALC sets different levels of error bounds, and also sets a threshold for the minimal compression ratio. When the compression ratio is great than the threshold, AALC uses the lowest error bound (lossless compression) for the state vector compression. If the compression ratio is less than the threshold, AALC re-compresses the state vector again with the next level of error bound. 

\begin{algorithm}[htbp]
 \caption{Quantum circuit simulation with AALC.}
 \label{algo:aalc}
\begin{algorithmic}[1]
\STATE $\Delta=\delta_0, \delta_1, \delta_2, ..., \delta_n$
\STATE $\theta$: compression ratio threshold
 \FOR {gate $U_i$ in the program}
   \FOR{stride $s_j$ in the state vector}
   \STATE decompress($s_j$)\\
   \STATE normalize($s_j$)\\
   \STATE gateComputation($U_i, s_j$)\\
   \FOR{$\delta_k$ in $\delta_0, ..., \delta_n$}
     \STATE ratio = compress($\delta_k, s_j$)\\
     \IF{ratio $\geq \theta$} 
     \STATE break
     \ENDIF
   \ENDFOR
  \ENDFOR
  \ENDFOR
\end{algorithmic}
\end{algorithm}

\section{Experimental Results} \label{results}

\begin{figure*}
    \centering
    \begin{subfigure}[!t]{0.48\textwidth}
        \includegraphics[width=\textwidth]{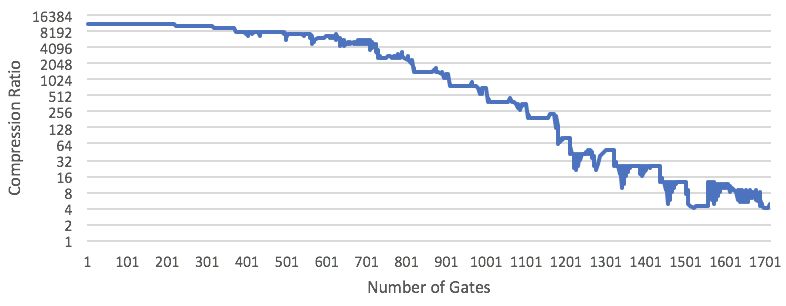}
        \caption{Threshold $\theta = 4$.}
        \label{fig:aalc_4}
    \end{subfigure}
    ~ 
    \begin{subfigure}[!t]{0.48\textwidth}
        \includegraphics[width=\textwidth]{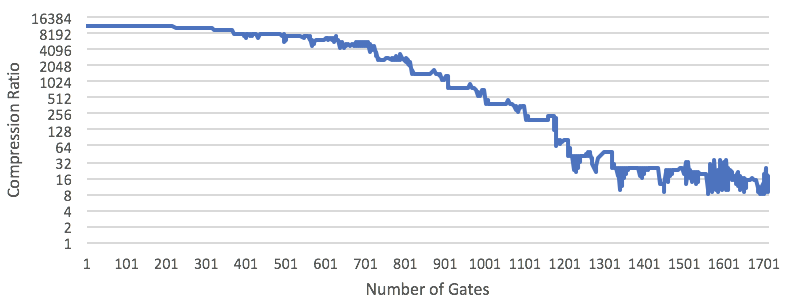}
        \caption{Threshold $\theta = 8$.}
        \label{fig:aalc_8}
    \end{subfigure}\hspace{5mm}
    ~
    \begin{subfigure}[!t]{0.48\textwidth}
        \includegraphics[width=\textwidth]{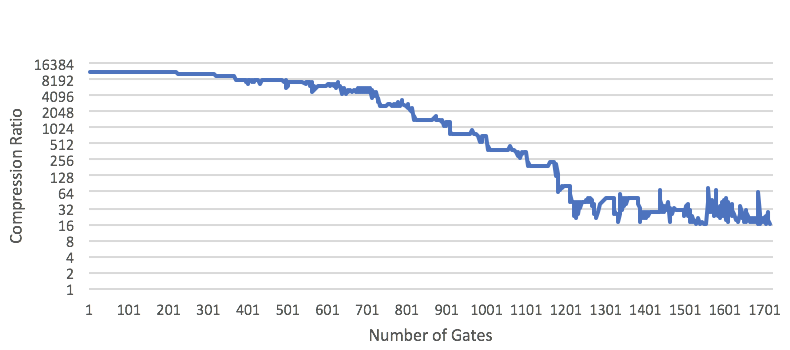}
        \caption{Threshold $\theta = 16$.}
        \label{fig:aalc_16}
    \end{subfigure}
    ~
    \begin{subfigure}[!t]{0.48\textwidth}
        \includegraphics[width=\textwidth]{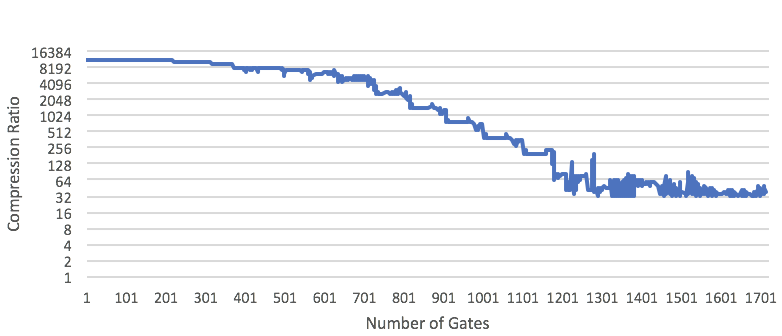}
        \caption{Threshold $\theta = 32$.}
        \label{fig:aalc_32}
    \end{subfigure}
    \caption{Compression ratio results of QFT quantum circuit simulation with AALC.}\label{fig:aalc_ratios}
\end{figure*}

To test the effectiveness of our AALC, we construct a set of error bounds to compress the state vector efficiently and perform the simulation with different compression ratio thresholds. We report the experimental results of QFT. QFT is a fundamental function of several quantum algorithms \cite{shor1999polynomial, kitaev1995quantum, mosca1999hidden, ettinger2004quantum}.  Figure~\ref{fig:aalc_ratios} shows the compression results with compression ratio threshold $\theta=4, 8, 16, \text{and } 32$. All the compression ratios of each simulation are above the threshold.

\begin{figure}[!t]
\centering
\captionsetup{justification=centering}
\includegraphics[width=0.48\textwidth,keepaspectratio]{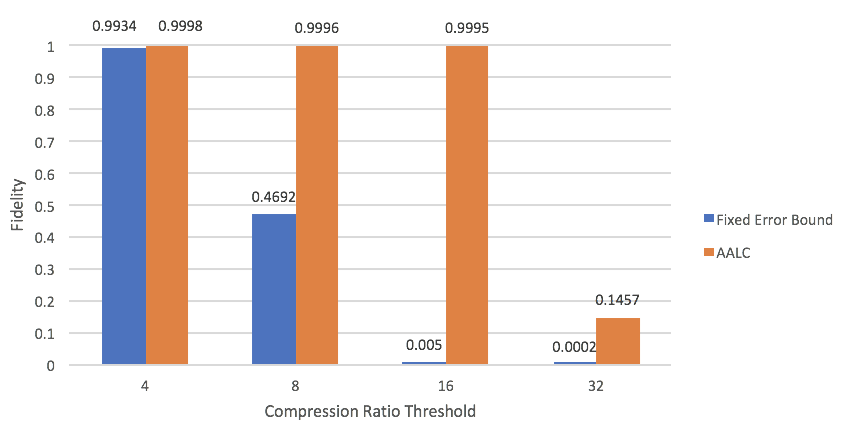}
\caption{Fidelity results of QFT quantum circuit simulation.}
\label{fig:fidelity}
\end{figure}

Figure~\ref{fig:fidelity} shows the final state fidelity results with AALC and without AALC. When we only use the fixed error bound for the lossy compression approach, the fidelity is 99.34\% when the compression ratio is above 4. However, the fidelity drops to 46.92\% when we use a larger error bound such that the compression ratio is above 8. The fidelity is only 0.05\% when the compression ratio threshold is 16. With our AALC technique, the fidelity results are 99.98\% and 99.96\% for the thresholds $\theta=4 \text{ and } \theta = 8$, respectively. The most significant result is that while the compression threshold is 16, the state fidelity is 99.95\%. When the threshold is 32, the fidelity drops to 14.57\%. This is because the error bound is too large such that the important quantum information is not maintained properly in the state amplitude vector.

\begin{figure}[!t]
\centering
\captionsetup{justification=centering}
\includegraphics[width=0.48\textwidth,keepaspectratio]{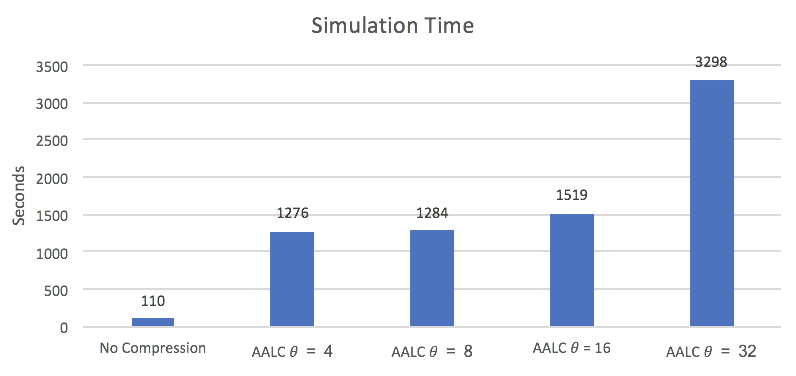}
\caption{Simulation run-time of QFT quantum circuit simulation.}
\label{fig:time}
\end{figure}

The simulation time is shown in Figure~\ref{fig:time}. Since our approach introduces the processes of data compression and decompression into the simulation, the run-time overhead comparing between no compression and AALC is from 11 to 32 times. For AALC, the higher threshold needs more time to complete the simulation because it has to test several error bounds to find the suitable one to exceed the threshold.

\begin{figure}[!t]
\centering
\captionsetup{justification=centering}
\includegraphics[width=0.48\textwidth,keepaspectratio]{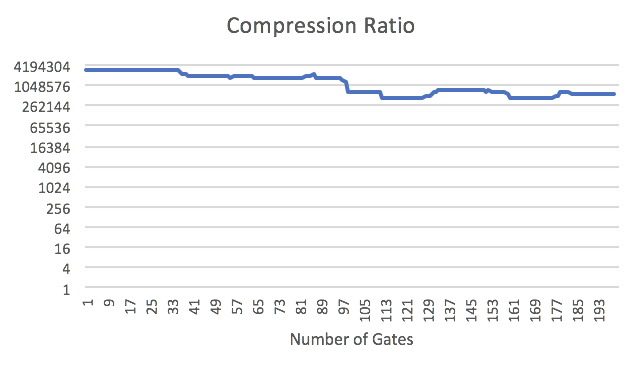}
\caption{Compression ratio results of Grover's search algorithm simulation with AALC.}
\label{fig:grover}
\end{figure}

QFT generates a lot of non-zero amplitudes, and there is no rule for the distribution of the amplitudes. Hence, amplitudes generated by QFT are hard to compress. QFT is a worst-case scenario for our approach. Let us see the effectiveness of AALC on another quantum algorithm. Grover’s search algorithm is one of the most famous quantum algorithms. In the simulation of Grover’s search algorithm,  most  of  the  state  amplitudes  are the same value. Such  state vectors allow the our AALC approach to perform high compression ratios. In our Grover’s search benchmark (30 qubits), the minimum compression ratio is 445144 (Figure~\ref{fig:grover}). The state fidelity is 99.75\%, and the simulation run-time overhead is 19 times in this experiment. This result suggests that using compression technique in some specific quantum circuit simulation may give us the chance to increase the simulation size significantly. For example, the compression, 445144, allows us to increase 18 ($\lfloor{log_2445144}\rfloor$) qubits for Grover's search algorithms simulation.

\section{Discussion and Future Directions} \label{future}

For general-purpose circuit simulation, AALC is able to reduce the data size to 1/16. The  reduction of memory requirement suggests that we could increase the size of the quantum circuit  simulation by 4 qubits comparing  to  the simulation without our AALC approach. For some  specific quantum circuits, like Grover’s search, our experimental results show that we could increase the simulation size by 18 qubits.  In the previous work, it is impossible for full amplitude vector update simulators  to  validate  the  quantum  algorithms  using  more than 49  qubits. In this paper, we propose a quantum circuits simulation technique to simulate more qubits than previously reported by using amplitude-aware lossy data compression. We trade computation power for memory space, and we further trade data accuracy for higher compression ratio.

Our ongoing work is to improve the integration of AALC and Intel-QS running on Argonne supercomputer system, Theta. Theta has 0.8 petabytes memory and is able to simulate a 45-qubit quantum computation. With our AALC approach, we expect to run 63-qubit Grover's search algorithm simulation and 49-qubit general quantum algorithms.

To further improve the simulation size, we need to maximize the compression ratio. It is essential to build a new type of compression algorithm for quantum state amplitudes. We plan to further improve AALC by introducing new type of lossy compression algorithm. This may take more steps to complete the compression and decompression, but we will be able to simulate the quantum circuit that we originally cannot simulate.

The run-time performance of AALC is limited by the iteration of searching for the proper error bound. If we have a good prediction model, then the process of trying different error bounds can be reduced significantly. In fact, the gate operation provides a prediction direction. For example, X, CNOT, and toffoli gates do not change the value of the amplitudes, and they only change the order of the amplitudes. Since this permutation does not affect the compression ratio significantly, we could safely use the error bound in the previous cycle for the current state vector to get similar compression ratio. In this way, we can save time to search the error bound. In addition, we do not need to compress every stride if the compression ratio is much higher than the compression threshold. In this situation, only a portion of the state vector needs to be compressed to fit in the memory requirement. There are a lot of opportunities for us to further improve the run-time performance of AALC. 

The enhanced error bound selection algorithm not only improves the run-time performance but also increase the state fidelity substantially. From the current compression ratio results, we notice that sometime the compression ratio is ``too good''. This means the selected error bound is too large, and we can actually choose a smaller error bound such that the compression ratio is still above the threshold. If there exists an oracle that always returns the smallest error bound which allows the compression ratio to be still above the threshold, then we can minimize the error impact on the state fidelity.

\section{Acknowledgments}
This research used resources of the Argonne Leadership Computing Facility, which is a DOE Office of Science User Facility supported under Contract DE-AC02-06CH11357.
This research was supported by the Exascale Computing Project (ECP), Project Number: 17-SC-20-SC, a collaborative effort of two DOE organizations - the Office of Science and the National Nuclear Security Administration, responsible for the planning and preparation of a capable exascale ecosystem, including software, applications, hardware, advanced system engineering and early testbed platforms, to support the nation's exascale computing imperative. The material was supported by the U.S. Department of Energy, Office of Science, and supported by the National Science Foundation under Grant No. 1619253. 
This work is funded in part by EPiQC, an NSF Expedition in Computing, under grant CCF-1730449. This work is also funded in part by NSF PHY-1818914 and a research gift from Intel.

\bibliographystyle{./IEEEtran}
\bibliography{./refs}

\begin{thebibliography}{10}
\providecommand{\url}[1]{#1}
\csname url@samestyle\endcsname
\providecommand{\newblock}{\relax}
\providecommand{\bibinfo}[2]{#2}
\providecommand{\BIBentrySTDinterwordspacing}{\spaceskip=0pt\relax}
\providecommand{\BIBentryALTinterwordstretchfactor}{4}
\providecommand{\BIBentryALTinterwordspacing}{\spaceskip=\fontdimen2\font plus
\BIBentryALTinterwordstretchfactor\fontdimen3\font minus
  \fontdimen4\font\relax}
\providecommand{\BIBforeignlanguage}[2]{{%
\expandafter\ifx\csname l@#1\endcsname\relax
\typeout{** WARNING: IEEEtran.bst: No hyphenation pattern has been}%
\typeout{** loaded for the language `#1'. Using the pattern for}%
\typeout{** the default language instead.}%
\else
\language=\csname l@#1\endcsname
\fi
#2}}
\providecommand{\BIBdecl}{\relax}
\BIBdecl

\bibitem{de2007massively}
K.~De~Raedt, K.~Michielsen, H.~De~Raedt, B.~Trieu, G.~Arnold, M.~Richter,
  T.~Lippert, H.~Watanabe, and N.~Ito, ``Massively parallel quantum computer
  simulator,'' \emph{Computer Physics Communications}, vol. 176, no.~2, pp.
  121--136, 2007.

\bibitem{smelyanskiy2016qhipster}
M.~Smelyanskiy, N.~P. Sawaya, and A.~Aspuru-Guzik, ``qhipster: the quantum high
  performance software testing environment,'' \emph{arXiv preprint
  arXiv:1601.07195}, 2016.

\bibitem{haner20170}
T.~H{\"a}ner and D.~S. Steiger, ``0.5 petabyte simulation of a 45-qubit quantum
  circuit,'' in \emph{Proceedings of the International Conference for High
  Performance Computing, Networking, Storage and Analysis}.\hskip 1em plus
  0.5em minus 0.4em\relax ACM, 2017, p.~33.

\bibitem{bernstein1997quantum}
E.~Bernstein and U.~Vazirani, ``Quantum complexity theory,'' \emph{SIAM Journal
  on computing}, vol.~26, no.~5, pp. 1411--1473, 1997.

\bibitem{markov2008simulating}
I.~L. Markov and Y.~Shi, ``Simulating quantum computation by contracting tensor
  networks,'' \emph{SIAM Journal on Computing}, vol.~38, no.~3, pp. 963--981,
  2008.

\bibitem{pednault2017breaking}
E.~Pednault, J.~A. Gunnels, G.~Nannicini, L.~Horesh, T.~Magerlein,
  E.~Solomonik, and R.~Wisnieff, ``Breaking the 49-qubit barrier in the
  simulation of quantum circuits,'' \emph{arXiv preprint arXiv:1710.05867},
  2017.

\bibitem{boixo2017simulation}
S.~Boixo, S.~V. Isakov, V.~N. Smelyanskiy, and H.~Neven, ``Simulation of
  low-depth quantum circuits as complex undirected graphical models,''
  \emph{arXiv preprint arXiv:1712.05384}, 2017.

\bibitem{chen201864}
Z.-Y. Chen, Q.~Zhou, C.~Xue, X.~Yang, G.-C. Guo, and G.-P. Guo, ``64-qubit
  quantum circuit simulation,'' \emph{Science Bulletin}, 2018.

\bibitem{chen2018classical}
J.~Chen, F.~Zhang, M.~Chen, C.~Huang, M.~Newman, and Y.~Shi, ``Classical
  simulation of intermediate-size quantum circuits,'' \emph{arXiv preprint
  arXiv:1805.01450}, 2018.

\bibitem{nielsen2002quantum}
M.~A. Nielsen and I.~Chuang, ``Quantum computation and quantum information,''
  2002.

\bibitem{zstd}
Y.~Collet, ``Zstandard - real-time data compression algorithm,''
  \emph{http://facebook.github.io/zstd/}, 2015.

\bibitem{xin2018}
D.~T. Z. C. F.~C. Xin~Liang, Sheng~Di, ``Efficient transformation scheme for
  lossy data compression with point-wise relative error bound,'' in
  \emph{CLUSTER}.\hskip 1em plus 0.5em minus 0.4em\relax IEEE, 2018.

\bibitem{tao2017significantly}
D.~Tao, S.~Di, Z.~Chen, and F.~Cappello, ``Significantly improving lossy
  compression for scientific data sets based on multidimensional prediction and
  error-controlled quantization,'' in \emph{Parallel and Distributed Processing
  Symposium (IPDPS), 2017 IEEE International}.\hskip 1em plus 0.5em minus
  0.4em\relax IEEE, 2017, pp. 1129--1139.

\bibitem{di2016fast}
S.~Di and F.~Cappello, ``Fast error-bounded lossy hpc data compression with
  sz,'' in \emph{2016 IEEE International Parallel and Distributed Processing
  Symposium (IPDPS)}.\hskip 1em plus 0.5em minus 0.4em\relax IEEE, 2016, pp.
  730--739.

\bibitem{divincenzo1995two}
D.~P. DiVincenzo, ``Two-bit gates are universal for quantum computation,''
  \emph{Physical Review A}, vol.~51, no.~2, p. 1015, 1995.

\bibitem{shor1999polynomial}
P.~W. Shor, ``Polynomial-time algorithms for prime factorization and discrete
  logarithms on a quantum computer,'' \emph{SIAM review}, vol.~41, no.~2, pp.
  303--332, 1999.

\bibitem{kitaev1995quantum}
A.~Y. Kitaev, ``Quantum measurements and the abelian stabilizer problem,''
  \emph{arXiv preprint quant-ph/9511026}, 1995.

\bibitem{mosca1999hidden}
M.~Mosca and A.~Ekert, ``The hidden subgroup problem and eigenvalue estimation
  on a quantum computer,'' in \emph{Quantum Computing and Quantum
  Communications}.\hskip 1em plus 0.5em minus 0.4em\relax Springer, 1999, pp.
  174--188.

\bibitem{ettinger2004quantum}
M.~Ettinger, P.~Hoyer, and E.~Knill, ``The quantum query complexity of the
  hidden subgroup problem is polynomial,'' \emph{arXiv preprint
  quant-ph/0401083}, 2004.

\end{thebibliography}

\end{document}